\documentclass[11pt]{article}
\usepackage{amsmath,amssymb}
\usepackage{epsfig,xspace,color,subfigure}

\newcommand{\nwc}{\newcommand}
\nwc{\Levy}{L\'evy}
\nwc{\LK}{L\'evy-Khintchine}
\nwc{\LI}{L\'evy-It\^{o}}
\nwc{\be}{\begin{equation}}
\nwc{\ee}{\end{equation}}
\nwc{\ba}{\begin{eqnarray}}
\nwc{\ea}{\end{eqnarray}}
\nwc{\nn}{\nonumber}
\nwc{\Z}{\mathbb{Z}}
\nwc{\CC}{\mathbb{C}}
\nwc{\R}{\mathbb{R}}
\nwc{\Rplus}{\R_+}
\nwc{\N}{\mathbb{N}}
\nwc{\M}{\mathcal{M}}
\nwc{\law}{\stackrel{\mathcal{L}}{\rightarrow}}
\nwc{\eqd}{\stackrel{\mathcal{L}}{=}}
\nwc{\ve}{\varepsilon} 
\nwc{\vp}{\varphi}
\nwc{\veps}{\varepsilon}
\nwc{\qref}[1]{(\ref{#1})}
\nwc{\D}{\partial}
\nwc{\dnto}{\downarrow}
\nwc{\LL}{\mathcal{L}}
\nwc{\nksup}{^{(n_k)}}
\nwc{\inv}{^{-1}}
\nwc{\one}{\mathbf{1}}

\nwc{\gena}{\mathcal{A}}
\nwc{\genb}{\mathcal{B}}
\nwc{\testfn}{\varphi}
\nwc{\jumpex}{n_*}
\nwc{\jump}{N}
\nwc{\Jbar}{j}
\nwc{\Kbar}{k}
\nwc{\drift}{b}
\nwc{\den}{F}
\nwc{\velu}{V_y}
\nwc{\velv}{V_z}
\nwc{\genadag}{\gena^\dagger}
\nwc{\genbdag}{\genb^\dagger} 
\nwc{\jumpsmol}{n}
\newcommand{\ignore}[1]{}

\begin{document}

\title{Coarse graining, dynamic renormalization and the kinetic theory of shock clustering}

\author{Xingjie Li \thanks{Division of Applied Mathematics, Brown University, 182 George St., Providence, RI 02912, USA. Email: {\tt xingjie\_li@brown.edu}. },
\and Matthew O. Williams \thanks{Program in Applied and Computational Mathematics, Princeton University, Fine Hall, Washington Rd., Princeton, NJ 08544, USA. Email:  {\tt mow2@princeton.edu}. },
\and Ioannis G. Kevrekidis \thanks{Department of Chemical and Biological Engineering, Princeton University, 6 Olden St.,  Princeton, NJ 08544, USA. Email:  {\tt yannis@princeton.edu}.},
\and Govind Menon \thanks{Division of Applied Mathematics, Brown University, 182 George St., Providence, RI 02912, USA, Email: {\tt govind\_menon@brown.edu}.}}

\maketitle


\vspace{2pc}
\noindent{\it Keywords\/}: Dynamic scaling, equation free method, Smoluchowski's coagulation equation, sticky particles, Burgers turbulence.

\begin{abstract}
{ 
We demonstrate the utility of the equation-free methodology developed by one of the authors (I.G.K) for the study of scalar conservation laws with disordered initial conditions. The numerical scheme is benchmarked on  exact solutions in Burgers turbulence corresponding to \Levy\/ process initial data. For these initial data, the kinetics of shock clustering is described by Smoluchowski's coagulation equation with additive kernel. The equation-free methodology is used to develop a particle scheme that computes self-similar solutions to the coagulation equation, including those with fat tails.   
}
\end{abstract}

\section{Introduction}
\label{sec:intro} 
We  combine two recent advances in this paper-- (a) the development of {\em equation-free numerical schemes\/} for multiscale problems~\cite{YK1,YK2}; and (b) the development of a {\em kinetic theory for shock clustering\/} in scalar conservation laws with random initial data~\cite{M1,MS1,MP4}. The essence of the equation free method is to extract the evolution of coarse macroscopic statistics for a system of  microscopically evolving particles by designing many brief parallel ``bursts'' of short-time evolution for the  microscopic system. Equation-free schemes are of most value when the microscopic evolution is fast and complex (given for example, by  a detailed, but expensive,  multiphysics code),  but the evolution of macroscopic variables is  slow and their evolution equations unknown. The fact that the closed evolution equations for the macroscopic statistics is unknown, or not known in closed form, is what makes these methods ``equation-free''. Nevertheless, as in all numerical methods, it is important to validate these schemes on model systems that are reasonably complex, but for which closed form equations for the coarse-grained problem are available.

The work presented here bridges this gap. We focus on the macroscopic statistics of the entropy solution to scalar conservation laws  with random initial data. To fix ideas, consider {\em Burgers turbulence\/}: the problem of determining the statistics of the solution to Burgers equation with a random velocity field, such as Brownian motion or white noise. The velocity field immediately develops infinitely many shocks separated by steep rarefaction waves, which cluster and decay as time increases. As one may expect, the process of shock clustering is complex (Burgers was motivated by  turbulence~\cite{Burgers}). Nevertheless, for certain classes of random data (including Brownian motion and white noise), the evolution of shock statistics is closed, and in fact, {\em exactly solvable\/}. In recent work, one of the authors (G.M.) and R. Srinivasan, derived kinetic equations that describe the clustering of shocks for any  scalar conservation with convex flux $f$, and random initial data within a large class~\cite{MS1}. Burgers turbulence is an interesting, but particular, instance of this theory.

The combination of the equation-free method and the kinetic theory of shock clustering can now be explained. Each microscopic state here is a spatial random field -- the random velocity field $u(x,t)_{x \in \R}$ at any instant in time, and the microscopic interaction is the rapid clustering of many shocks in a short time frame. The macroscopic statistics are the probability distribution of $u(x,t)_{x \in \R}$  (its $n$-point distribution functions). We compare the  statistics computed via the equation-free scheme with the exact solutions given by the kinetic theory. 

Our aims in this work are two-fold: (a) to demonstrate the utility of the equation-free methodology for computing dynamic scaling in shock clustering; (b) to present the exact solutions in shock clustering as a useful benchmark problem for other practitioners in multiscale methods. For these reasons, this paper is organized as follows. We first review the exact solubility of scalar conservation laws with Markov process data, and the kinetic theory of shock clustering in some detail. We then turn to the interpretation of these systems within the equation-free methodology. Finally, we turn to a set of numerical experiments that illustrate the method on a basic test case: the statistics of shocks to Burgers equation with \Levy\/ process initial data. In this case, the kinetic equations of~\cite{MS1} reduce to a basic model of clustering -- Smoluchowski's coagulation equation with additive kernel. The equation free method provides a new numerical scheme for Smoluchowski's coagulation equation. This method is shown to accurately and efficiently compute all self-similar solutions, including those with fat tails.

\section{Background}
\label{sec:bg}
\subsection{Resolving the closure problem}
One of the central obstructions in studies of turbulence  (e.g. in homogeneous isotropic turbulence in incompressible fluids) is the closure problem: the evolution equations for $n$-point statistics involve  $n+1$-point statistics. The results presented in~\cite{MS1} resolve the closure problem for a tractable, but fundamental, class of nonlinear partial differential equations. Consider a scalar conservation law on the line
\ba
\label{eq:sc}
&&  \partial_t u + \partial_x f(u) =0, \quad -\infty < x < \infty, \quad t >0,\\
\label{eq:sc2}
&& u(x,0) = u_0(x),
\ea
with a convex, $C^1$ flux $f$. The unique entropy solution to \qref{eq:sc} is given by the Hopf-Lax formula (e.g.~\cite[\S 1.1]{MS1}). The two main results in~\cite{MS1} are as follows:
\begin{enumerate}
\item {\em Closure theorem:\/} If $u_0(x)$ is a Markov process (in $x$) with only downward jumps (a {\em spectrally negative Markov process\/}), then so is the solution $u(x,t)$ for each $t>0$.
\item {\em Kinetic theory:\/} The infinitesimal generator of $u(x,t)$ satisfies a Lax equation (equation~\qref{eq:kinetic} below) that describes the kinetics of shock clustering.       
\end{enumerate}
The closure theorem shows that a large class of random processes is left invariant by the Hopf-Lax formula.  Since the $n$-point function for a Markov process on the line factors into $1$ and $2$-point distribution functions, the closure theorem tells us that the evolution of these functions determines the evolution of $n$-point statistics exactly. The generator provides an efficient representation of $2$-point statistics: informally, it is the derivative of the $2$-point distribution function as the gap between the $2$-points shrinks to zero. It is simplest to explain its form under the assumption that $u(x,t)$ is a stationary  Markov process (in $x$) with mean zero. In this case, for each $t>0$, the generator $\gena(t)$ is an integro-differential operator  that acts on test functions $\testfn \in C_c^{1}(\R)$ via
\be
\label{eq:gena1}
\gena(t)\testfn(u) = \drift(u,t) \testfn'(u) + \int_{-\infty}^u \left( \testfn(v)-\testfn(u)\right) \jump(u,dv,t). 
\ee
The jump kernel $\jump(u,dv,t)$ describes the rate of jumps (shocks) from state $u$ to state $v$ at time $t$. Observe that the velocity field $u(x,t)$ jumps only downwards as $x$ increases (i.e. $u >v$). However, this does not mean that $u(x,t)$, $x \in \R$ is decreasing -- it can increase continuously through rarefactions -- this is described by the drift coefficient $\drift(u,t)$. 

We use the flux function $f$ and the drift and jump measure of $\gena$ to define a second operator
\be
\label{eq:genb}
\genb \testfn (u) = -f'(u)\drift(u,t) \testfn'(u) -\int_{-\infty}^u \frac{f(u)-f(v)}{u-v }\left( \testfn(v)-\testfn(u)\right) \jump(u,dv,t). 
\ee
Then the Lax equation derived in~\cite{MS1} is
\be
\label{eq:kinetic}
\partial_t \gena = [\gena, \genb] = \gena \genb -\genb \gena.
\ee
The compact form of \qref{eq:kinetic} is equivalent to (lengthy, but intuitive) Vlasov-Boltzmann  equations for the drift $b(u,t)$ and jump kernel $\jump(u,dv,t)$ obtained by  substituting the definitions  \qref{eq:gena1}--\qref{eq:genb} in the Lax equation \qref{eq:kinetic} (see~\cite[equations (26)--(30)]{MS1}). These are the kinetic equations for shock clustering.

\subsection{An equation free approach to shock clustering} 
\label{subsec:eq-setup}
The equation-free methodology is applicable to systems with evolution on two decoupled scales -- fast evolution of microscopic states and slow evolution for macroscopic statistics that describe averages over the microscopic states. The evolution of the microscopic states is assumed to be known. The evolution of macroscopic statistics is assumed to satisfy a closed equation, but the precise form of this equation is not assumed to be known, and is computationally approximated via a {\em coarse evolver\/} as follows. The macroscopic statistic at time $t$ is  (i) ``lifted'' into an ensemble of  microscopic states consistent with this macroscopic statistic; (ii) each microscopic state in the ensemble is evolved by the fast evolution over a time step $\Delta t$; (iii) the macroscopic statistic at time $t+\Delta$ is obtained by averaging over the ensemble of microscopic states at time $t+\Delta t$.

We now combine the kinetic theory of shock clustering with the equation-free methodology. Assume $t>0$ is fixed. A microscopic state is a spatial random field $u(x,t)_{x \in \R}$. The microscopic evolution is the clustering of shocks and the decay of rarefactions.  The macroscopic statistics are its  $1$ and $2$-point functions. Since the $1$ and $2$-point functions  can be computed once $\gena(t)$ is known, an equivalent macroscopic statistic is the generator $\gena(t)$, and the closed macroscopic evolution is given by the Lax equation \qref{eq:kinetic}. This (exact) evolution is contrasted with the computational coarse evolver that uses only the microscopic evolution of shocks and rarefactions. 
 
Thus, for this particular application, the coarse evolver of the equation-free scheme consists of three steps:
\begin{enumerate}
\item Sample $P$ realizations of the Markov process $u(x,0)$ given its generator $\gena(0)$. Call these $u_j(x,0)$, $j =1, \ldots, P$.
\item Evolve each realization $u_j$ in parallel for a short burst of time $\Delta t$ by the PDE~\qref{eq:sc}. This has a simple particle interpretation -- the shocks behave like sticky particles -- with a rule of `stickiness' determined by $f$.  
\item Estimate the generator $\gena(\Delta t)$ from the $P$ realizations  $u_j(x,\Delta t)$, $j=1,\ldots, P$.  In practice, this is the most difficult step. 
\end{enumerate} 
At the end of the short time burst, $\Delta t$, we have progressed from  $\gena(0)$ to $\gena(\Delta t)$. In general, the time evolution of $\gena(t)$ may now be accelerated by using the difference $(\gena(\Delta t)-\gena(0))/\Delta t$ as an estimate of $\dot{\gena}$ at $t=0$. For example, this estimate can be fed into an Euler scheme with a  time-step $\Delta T \gg \Delta t$. 

In the examples treated in this paper, the shocks cluster into larger and larger shocks as time evolves, and the natural long-time limit to consider is {\em self-similar shock statistics\/}. We use two distinct techniques to accelerate the time evolution to capture the self-similar solutions.  
The first is {\em dynamic renormalization\/}. After time $\Delta t$ we suitably rescale $\gena(\Delta t)$ before using it as the input to the next step of the microscopic evolver. This approach can only be used to compute self-similar solutions that are stable (in rescaled variables). In the second approach, the self-similar solution is reformulated as a {\em fixed point problem\/}. Self-similar solutions are then determined via a Newton-GMRES scheme. The advantage of this approach is that the method will converge quadratically (given a sufficiently good initial guess) regardless of the stability of the desired self-similar solution. Both these approaches have been explored in previous work by one of the authors (I.G.K) and his co-workers (see e.g.~\cite{YKPZ}). The main novelty here lies in the application of these techniques to shock-clustering.  In order to describe the implementation of these ideas, we now describe some exact solutions to shock clustering in greater detail.

\section{Exact solutions: theory and computation}
\subsection{The Burgers-\Levy\/ case}
\label{subsec:exact}
The work~\cite{MS1} builds on two sets of results for Burgers equations: pioneering, but formal, calculations of Duchon and his students~\cite{Duchon1,CD}; and an important closure theorem of Bertoin~\cite{B_burgers}.  It is simplest to describe these results in the following situation. 

Consider the entropy solution to Burgers equation on the half-line $[0,\infty)$:
\ba
\label{eq:burgers}
&& \partial_t u + \partial_x \left( \frac{u^2}{2}\right)=0, \quad 0 < x < \infty, \; t>0\\
&& u(x,0) = u_0(x) \leq  0,
\ea
where $u_0(x)$ is a piecewise constant, decreasing \Levy\/ process. (A boundary condition at $0$ is not needed since characteristics only flow out of the domain $[0,\infty)$). In this context, Bertoin's closure theorem asserts that the process $u(x,t)-u(0,t)$,$x \geq 0$ remains a piecewise constant, decreasing \Levy\/ process for each $t>0$.  \Levy\/ processes are Markov processes with increments that are independent and identically distributed. Consequently, their jump kernel $\jump(u,dv)$ depends only on the difference $u-v$. By Bertoin's theorem, at any $t>0$, the generator $\gena(t)$ is  of the form~\footnote{We have assumed that the \Levy\/ measure of $u(x,t)$ has a density $f(s,t)$ for convenience. See~\cite{MP4,B_burgers} for the completely general statement}
\be       
\label{eq:gen_smol}
\gena(t)\testfn(u) = \int_0^\infty \left(\testfn(u-s)-\testfn(u) \right) f(s,t)\, ds. 
\ee
The general Vlasov-Boltzmann equation~\qref{eq:kinetic} now simplifies to {\em Smoluchowski's coagulation equation with additive kernel\/}:
\begin{equation}
\label{eq:smol}
\frac{\partial n}{\partial t}(s,t) = \frac{1}{2} \int_0^s s \, n(t,s-s')
n(s',t) ds'  - \int_0^\infty (s+s') \, n(s,t) n(s',t) ds',
\end{equation} 
where the number density $n(s,t)$ is related to the \Levy\/ density $f(s,t)$ by 
\be
\label{eq:levy-f}
n(s,t) = \frac{f(s,t)}{\int_0^\infty rf(r,t) \, dr}. 
\ee 

We briefly review an intuitive description of the link between~\qref{eq:burgers} and \qref{eq:smol}~\cite[\S 2.1]{MP4}. First, note that by restricting attention to piecewise constant, decreasing velocity fields, we have prevented the appearance of any rarefaction waves in the system. 
Let $m_0(t) = \int_0^\infty f(s,t) \, ds$ denote the expected number of jumps for the \Levy\/ process $u(x,t)$ in a unit interval and assume $m_0(0) < \infty$. Then  $m_0(t) \leq m_0(0) < \infty$ for each $t>0$ since the total number of shocks can only decrease by collisions.  
For each $t\geq 0$, the process $u(x,t)-u(0,t)$ with jump density $f(s,t)$ has the following form: 
\begin{enumerate}
\item The shock locations $0 =x_0(t) <  x_1(t) < x_2(t) < \ldots x_j(t) < \ldots $ form a Poisson process with rate $m_0(t)$. 
\item The size of the shocks $s_j(t)$ at the jump locations $x_j(t)$ are independent, identically distributed (iid) random variables  with probability density $ m_0(t)^{-1} f(s,t)$.
\item
The velocity difference $u(x,t)-u(0,t)$  is a piecewise constant function that takes the values
\be
\label{eq:vel-field}
u_k(x,t) = - \sum_{j=1}^{k-1} s_j, \quad x_{k-1} < x < x_k, \quad k \geq 1.
\ee
\end{enumerate}
In order that such a velocity field constitute a weak solution to \qref{eq:burgers}, the speed of shocks is given by the Rankine-Hugoniot relation 
\be
\dot{x}_k = - \sum_{j=1}^{k-1} s_j - \frac{s_k}{2}, 
\ee
When two shocks meet, they stick and the speed recomputed from the Rankine-Hugoniot relation with the new left and right limits. We compute the rate of growth and decay of individual shocks by summing over all possible collision events to obtain~\qref{eq:smol} (see~\cite[\S 2.1]{MP4} for details).

\subsection{Long time asymptotics}
The behavior of \qref{eq:smol} is  well understood~\cite{MP1}.  Consider the $p$th moment 
\be
\label{eq:moments-p}
M_p(t) = \int_0^\infty s^p n(s,t)\, ds,
\ee
and call $M_0(t)$ the {\em total number\/} and $M_1(t)$ the {\em total mass}~\footnote{This terminology is motivated by the origins of Smoluchowski's coagulation equation in physical chemistry~\cite{Drake}}. Then equation \qref{eq:smol} has a unique global solution for any initial measure with $M_1 <\infty$~\cite[Thm 2.8]{MP1} (other moments, including $M_0$ may be infinite). Further, the solution preserves mass, and  without loss of generality, we may rescale the initial data $n_0$ so that 
\be
\label{eq:moments}
M_1(t) = \int_0^\infty s n(s,t) \, ds = 1, \quad t \geq 0.
\ee
For each $\rho \in (0,1]$,  equation \qref{eq:smol} has a self-similar solution 
\begin{equation}\label{eq:nform}
n(s,t) = e^{-2t/\beta} n_\rho(e^{-t/\beta}s),
\end{equation}
where $\beta=\rho/(1+\rho)$,  and 
\begin{equation}
\label{eq:l6}
 n_\rho(s) = \frac1\pi \sum_{k=1}^\infty
 \frac{(-1)^{k-1}s^{k\beta-2}}{k!}
\Gamma(1+k-k\beta){\sin \pi k\beta}.
\end{equation} 
In the case $\rho=1$, the formula above simplifies to
\be
\label{eq:rho1}
n_1(s) =\frac{e^{-s/4}}{\sqrt{4\pi s^3}}.
\ee
Each self-similar solution has mass $1$. However, they differ in their asymptotics as $s \to \infty$. Only the solution for $\rho=1$ has an exponential tail; for each $0< \rho < 1$, we find the algebraic decay (``fat tail'')
\be
\label{eq:smol-tail}
n_\rho(s) \sim \frac{\rho +1}{|\Gamma(-\rho)|s^{-(2+\rho)}}
\quad s \rightarrow \infty.
\ee 
As a consequence, for any $0 < \rho < 1$, the $\rho+1$-st moment diverges logarithmically:
\be
\label{eq:log}
\int_0^s r^{1+\rho} n_\rho(r) \, dr  \sim \frac{\rho +1}{|\Gamma(-\rho)|} \log s, \quad  s \rightarrow \infty.
\ee 

All initial densities with $M_2 < \infty$ converge to the self-similar solution with $\rho=1$.  The approach to the fat-tailed self-similar solutions is delicate. Roughly speaking, an initial density $n(s,0)$ lies in the domain of attraction of $n_\rho$ if and only if 
the tails of $n(s,0)$ diverge in the same manner as \qref{eq:smol-tail} (see~\cite[Thm 7.1]{MP1} for necessary and sufficient conditions). This analytical subtlety is reflected in numerical calculations of self-similar solutions: a typical fixed point method for finding self-similar solutions usually converges to $n_1(x)$, not any of the fat-tailed solutions. Since the divergence in \qref{eq:log} is only logarithmic, we will impose the condition $M_{1+\rho} < \infty$ as a ``pinning condition'' in both the dynamic renormalization and Newton-GMRES schemes to compute the fat-tailed self-similar solutions $n_\rho$, $0 < \rho < 1$.

\subsection{Implementing the coarse evolver\/}
\label{subsec:coarse}
As described in Section~\ref{subsec:eq-setup}, implementation of the equation-free method requires an efficient scheme to estimate the jump kernel of a Markov process, given $P$ paths. This estimation problem is considerably simpler for the Burgers-\Levy\/ case. In order to understand the issue, imagine approximating the initial velocity field  $u(x,0)$ in \qref{eq:sc2}  by a Markov process with $M$ states $v_1 < \ldots < v_M$. In this case, the generator $A_M(0)$ is an $M \times M$ matrix and it is easy to sample $N$ velocity fields $u_j(x)$ generated by $A_M(0)$. Similarly, it is easy to evolve each random velocity field by \qref{eq:sc} using the Hopf-Lax formula, since a convex hull of $N$ points can be computed in $O(N\log N)$ steps. Thus, after time $\Delta t$ we have $P$ random velocity fields $u_j(x,\Delta t)$, and our task is to form the best estimate of the generator $A_M(\Delta t)$ from these samples. In  general, the matrix $A_M(\Delta t)$ has $O(M^2)$ terms. However, in the Burgers-\Levy\/ case, as a first approximation, the generator is a Toeplitz matrix with only $M$ terms. Thus, for fixed $M$, it can be estimated with higher accuracy even with relatively few realizations (smaller $P$). For these reasons, we focus on the Burgers-\Levy\/ case in this article. We expect to analyze the general Lax equation \qref{eq:kinetic} in future work.

We fix a maximal number of particles $N_0$ and a time step $\Delta t$.  The coarse evolver in our numerical computation takes the following form. 
\begin{enumerate}
	\item Assume given a \Levy\/ density $f(s,0)$ with $m_1(0)=1$ and $m_0(0) < \infty$.
	\item Generate the first $N_0$ jumps of a decreasing \Levy\/ process $u_0(x)$ with jump density $f(s,0)$. The initial length of the computational domain is $L(0)=x_{N_0}$.
	\item Evolve the \Levy\/ process by Burgers equation up to time $\Delta t$. This is done in one-step, either by the use of the Hopf-Cole formula, or by the sticky particle algorithm of~\cite{BG}. As noted above, this step involves the computation of a convex hull, and requires $O(N_0 \log N_0)$ steps (i.e. it is fast).
	\item Let $N(\Delta t)$ denote the number of particles in the system and let $L(\Delta t) = x_{N(\Delta t)}-x_1(\Delta t)$.  Compute the empirical \Levy\/ measure 
\be
\label{eq:empirical} 
f^{(e)}_e(s,\Delta t)\, ds = \frac{1}{L(\Delta t) N(\Delta t)}\sum_{k=1}^{N(\Delta t)} \delta_{s_k(\Delta t)}(ds).
\ee
\end{enumerate}

This is the coarse evolver for one trial. In fact,  $P$ trials can be run in parallel, and if the empirical \Levy\/ density of each of these is $f^{(e)}_j$, we further average over the $P$ trials to obtain the coarse evolution 
\be
\label{eq:empirical-ref} 
f^{(e)}(s,\Delta t)\, ds = \frac{1}{P}\sum_{j=1}^{P} f^{(e)}_j(s,\Delta t). 
\ee    
In practice, the scheme above has to be modified to streamline the computation. First,  we further smooth the empirical density in \qref{eq:empirical-ref} to simplify the task of sampling a \Levy\/ process with this empirical density when $f^{(e)}(\cdot,\Delta t)$ is used as input. Second, all the self-similar solutions have divergent total number (i.e. $\int_0^\infty n_\rho(s)\, ds=\infty$). The divergence arises from the number of small clusters (e.g. $n_1(s) \sim s^{-3/2}$ as $s \to 0$). At each step of the renormalization, the number $m_0(\Delta t)$ increases. The computation is terminated when $m_0$ crosses a fixed threshold (the maximal number we use is $2 \times 10^7$). We finally note that the \Levy\/ density \qref{eq:gen_smol} completely specifies the generator $\gena(t)$. Thus, we have demonstrated, as explained in Section~\ref{subsec:eq-setup}, that the coarse evolution is a map from $\gena(0)$ to $\gena (\Delta t)$.

\section{Numerical experiments}
\subsection{Fixed point equations}
\label{subsec:expts}
In the numerical experiments, we find it more convenient to work with the Smoluchowski density $n$, which is related to the \Levy\/ density $f$ through \qref{eq:levy-f}. It is helpful to denote the coarse evolver as follows: the procedure of Section~\ref{subsec:coarse} provides a map: $n \mapsto \mathcal{G}(n)$ for a Smoluchowski density $n(s)$ on $(0,\infty)$. This allows us to recognize the self-similar profiles as fixed points of a suitable map. Explicitly, we use \qref{eq:nform} to see that for each $\rho$, if $a_\rho = e^{2\Delta t/\beta}$ and $b_\rho=e^{\Delta t/\beta}$, with $\beta=\rho/(1+\rho)$ then
\be
n_\rho(s) := a_\rho \mathcal{G}(n_\rho(b_\rho s).
\ee 

These profiles are numerically computed as follows. We start by fixing a value for the parameter $\rho$ in the range $(0,1]$. Given a Smoluchowski density $n$ with compact support, let $\mathcal{R}_\rho (n)$ denote the rescaling of $n$ that satisfies the {\em pinning conditions\/}
\be
\label{eq:pinning}
\int_0^\infty s \mathcal{R}_\rho (n) (s) \, ds =1, \quad \int_0^\infty s^{1+\rho} \mathcal{R}_\rho (n) (s) \, ds =1.
\ee
For each $\rho \in (0,1]$ and a Smoluchowski density $n$ with sufficiently rapid decay, we define the renormalized mapping 
\be
\label{eq:grho}
\mathcal{H}_\rho(n) := \mathcal{R}_\rho \mathcal{G} \mathcal{R}_\rho.
\ee
The mapping $\mathcal{H}_\rho$ is a synthesis of time evolution and dynamic rescaling.  When $\rho=1$, the self-similar profile $n_1(s)$ is a fixed point of $\mathcal{H}_1$. For $0 < \rho < 1$,
it is {\em not\/} true that $n_\rho = \mathcal{H}_\rho(n_\rho)$. This is because $n_\rho$ does not have finite $1+\rho$-th moment. Nevertheless, this  moment is `critical' in terms of the asymptotic relation \qref{eq:log}, and the divergence is logarithmic. Thus, since we are restricted to a finite domain in computations, it is natural to seek the fat-tailed solutions as fixed points of $\mathcal{H}_\rho$.

We use two strategies to find the fixed point. The first is a direct iteration of the map above. We term this {\em dynamic renormalization\/}. The scheme is as follows. We first fix $0< \rho \leq 1$ and an initial Smoluchowski density $n^{(0)}$. We then generate a sequence of Smoluchowski densities via the iteration
\be
\label{eq:iterate}
n^{(k+1)} = \mathcal{H}_\rho \left( n^{(k)} \right).
\ee
A second method of solving this equation is to use a fixed point algorithm, such as the {\em Newton-GMRES} scheme. For any density $n$ we define the residual 
\[ r = n - \mathcal{H}_\rho(n)\]
and use a Newton iteration to solve this equation for $r=0$. In this setting, the combination of the Newton-Raphson method with the matrix-free GMRES scheme is particularly advantageous because the Jacobian, $\partial r/\partial n$ does not need to be computed explicitly.  Instead, a series of ``numerical experiments'' is used to approximate the Jacobian in a Krylov subspace.  In the results that will follow, the Newton iteration scheme is augmented with an Armijo line search to make the iteration scheme more robust to the choice of initial guess.
  
%
%

Note that neither  procedure selects $\rho$ automatically. Further, our choice of initial conditions is guided by $\rho$. We use a monodisperse initial condition for $\rho =1$ (all shocks of initial size $1$), and for other $\rho$ we choose the initial condition $n^{(0)}=s^{-(2+\rho)}$.  Thus, our  approach is certainly guided by {\em a priori\/} knowledge of the existence of a $1$-parameter family of self-similar solutions. In fact, earlier numerical schemes for the computation of self-similar solutions implicitly used the pinning condition $M_2=1$, and thus computer experiments did not reveal the existence of fat-tailed solutions~\cite{Lee}. We view this degeneracy as a useful cautionary note for the numerical computation of  self-similar solutions, here and in other problems.

\subsection{Results}
\label{subsec:results}
\begin{figure}
\centering
\subfigure[$\rho =1$ ]{\includegraphics[width =7.5 cm]{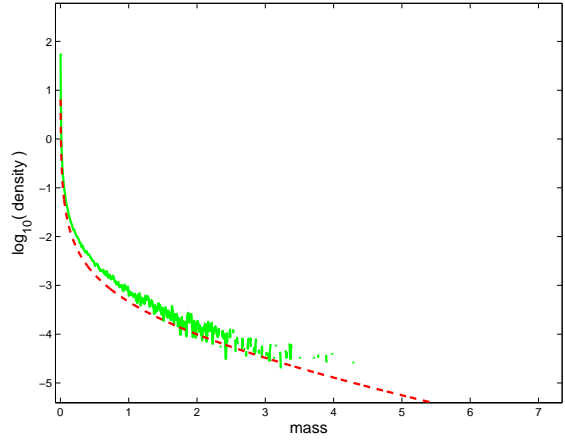}}
\subfigure[$\rho =0.5$]{\includegraphics[width =7.5 cm]{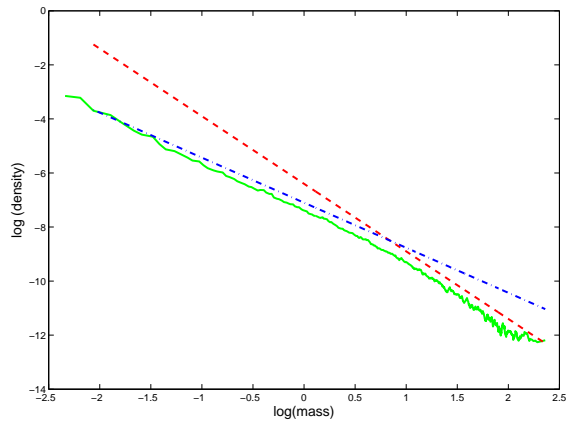}}
\caption{Density of exact and computed self-similar solutions for $\rho=1$ and $\rho=0.5$. The  lines in (b) correspond to rigorous asymptotics of $n_\rho$ as $s \to 0$ and $s \to \infty$.}
\label{fig:density1}
\end{figure}

Various representative results of our computations are presented here. In all the examples below, we denote the exact self-similar solution by $n_\rho$ and the numerically computed fixed point by $\tilde{n}_\rho$. We first compare the exact and computed densities  for $\rho=0.5$ (fat tails) and $\rho=1$ (exponential tails) in Figure~\ref{fig:density1}.
Since all densities are rescaled to have unit mass, we define the Kolmgorov-Smirnov statistic  between computed and exact results:
\[ d(n_\rho,\tilde{n}_\rho) = \sup_{s \geq 0} \left| F_\rho (s) -\tilde{F}_\rho(s) \right|, \]
where
\[ F_\rho(s) = \int_0^s r n_\rho(r) \, dr, \quad \tilde{F}_\rho(s) = \int_0^s r \tilde{n}_\rho (r) \, dr.\]
\begin{figure}
\centering
\subfigure[$F_1(s)$ and $\tilde{F}_1(s)$]{\includegraphics[width =7.5 cm]{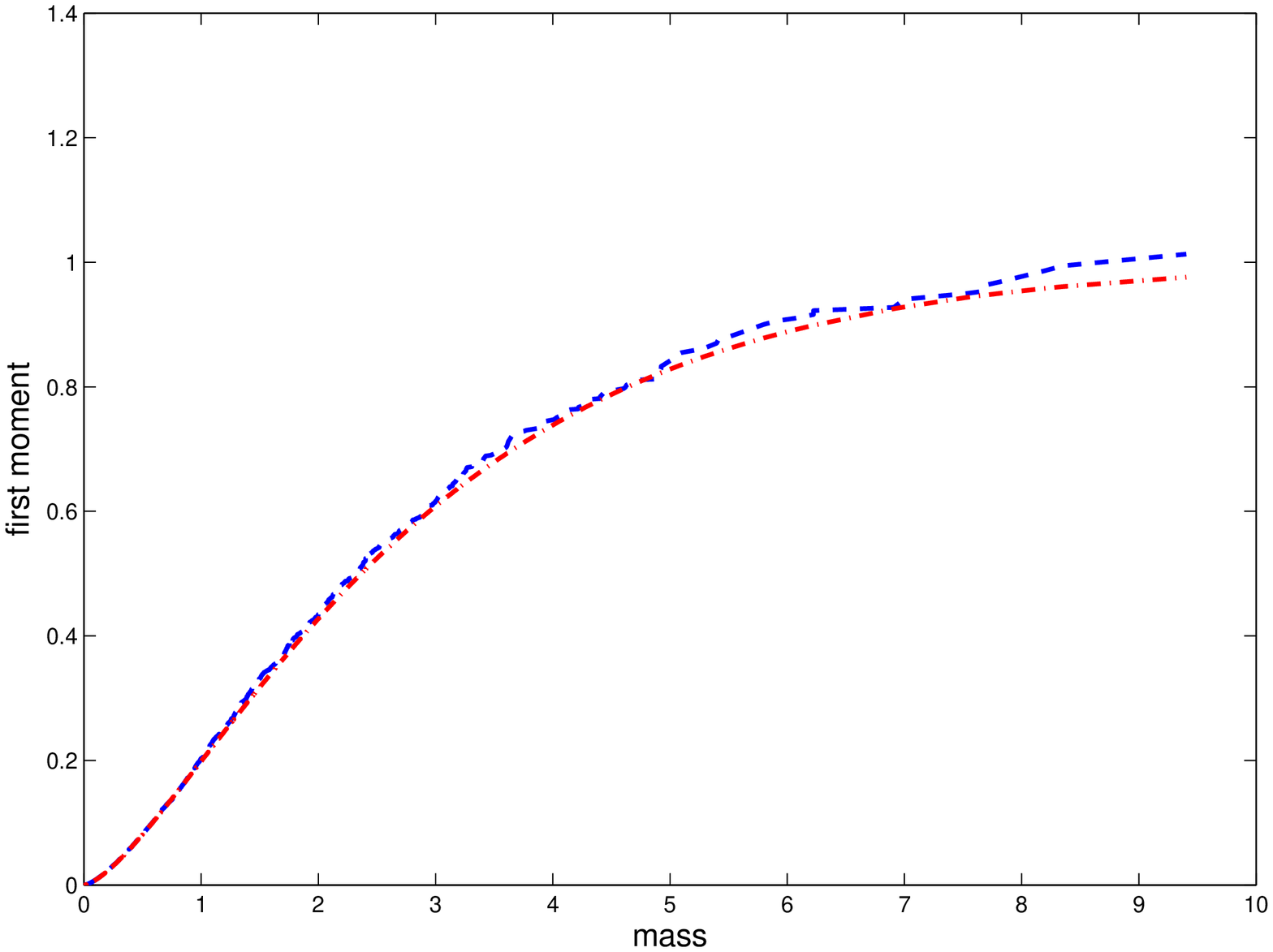}}
\subfigure[$|F_1 (s)-\tilde{F}_1(s)|$]{\includegraphics[width =7.5 cm]{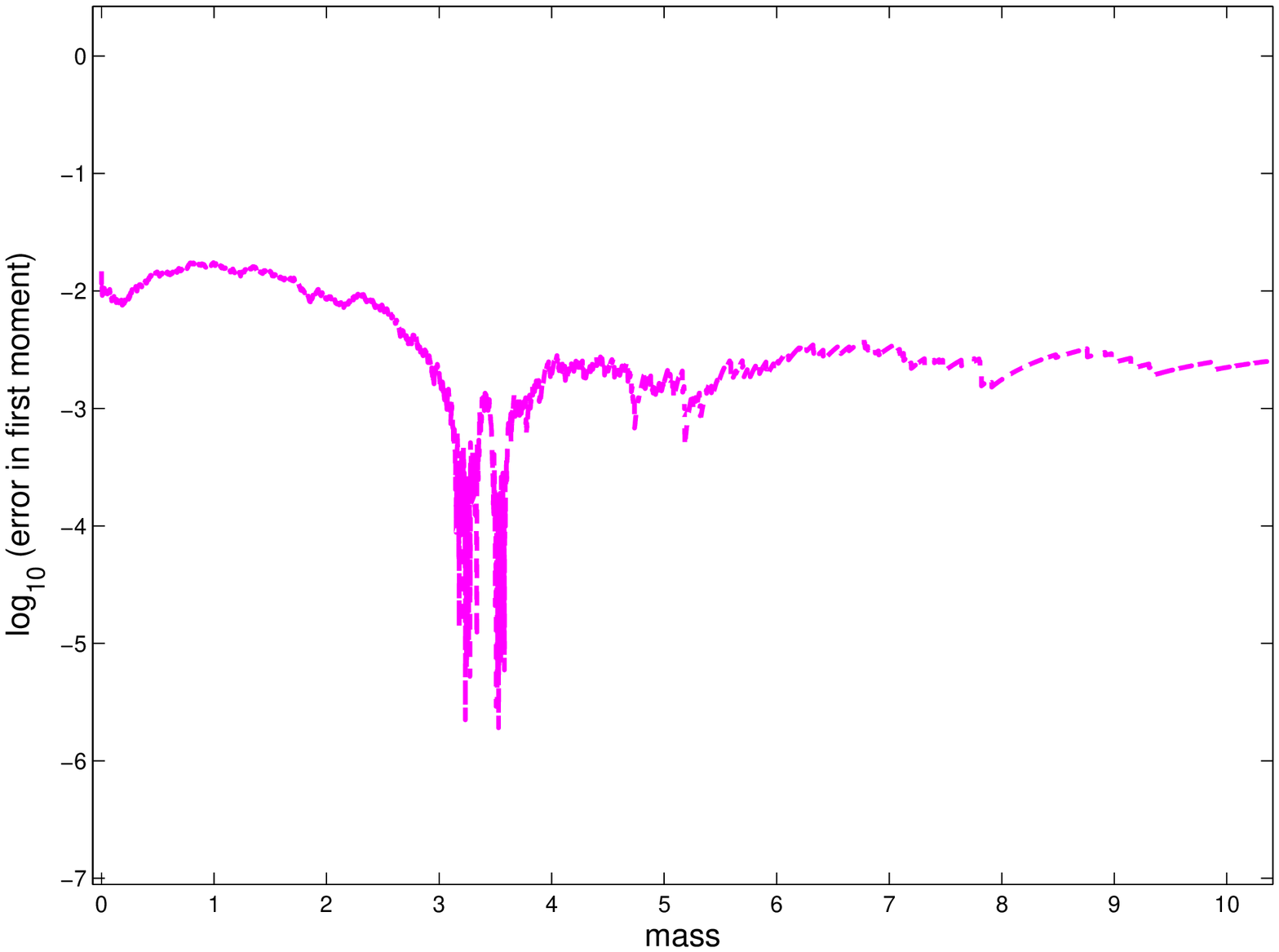}}
\caption{The empirical first moment and difference $|F_1(s)-\tilde{F}_1(s)|$ for $\rho=1$.}\label{fig:mono}
\end{figure}

\begin{figure}
\centering
\subfigure[Dynamic renormalization]{\includegraphics[width =8.2 cm, width =7.5 cm]{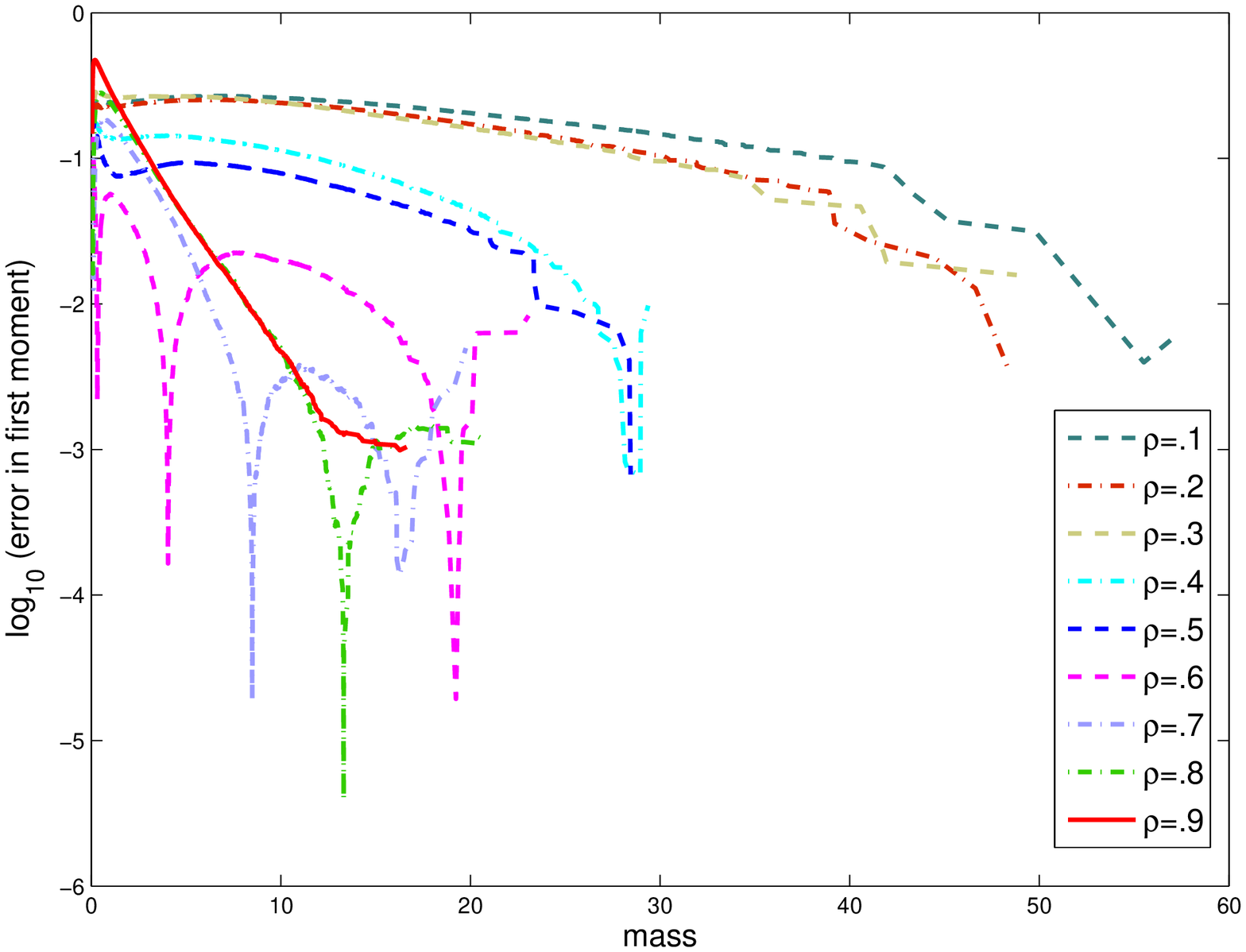}}
\subfigure[GMRES-Newton]{\includegraphics[width =8.2 cm, width =7.5 cm]{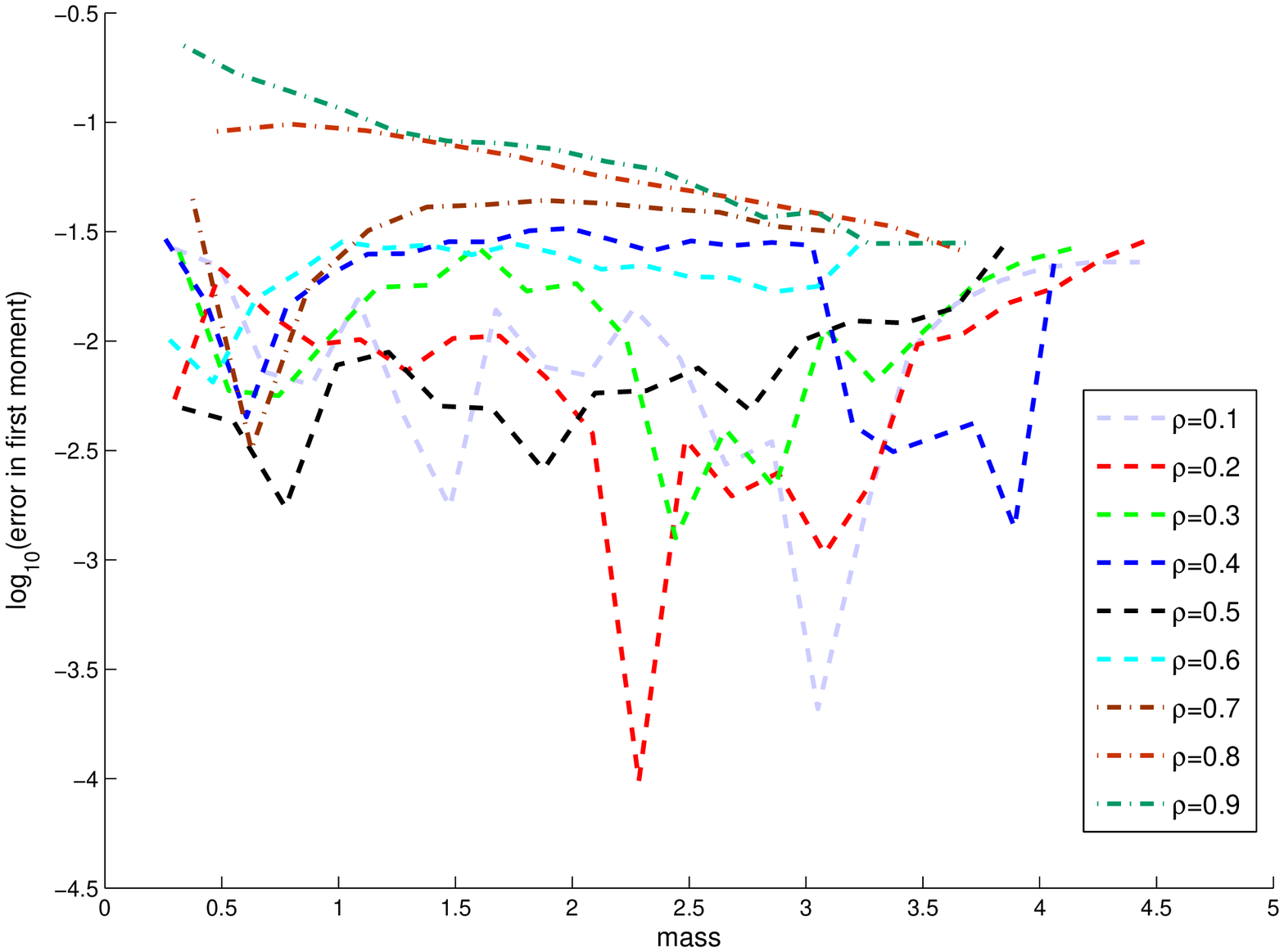}}
\caption{The sup-norm difference $|F_\rho(s) -\tilde{F}_\rho(s)|$ as a function of $s$ for $0<\rho<1$ using (a) dynamic renormalization; (b) Newton-GMRES.}\label{fig:fat}
\end{figure}
The comparison between $F_1$ and $\tilde{F}_1$ is shown in Figure~\ref{fig:mono}. Similar comparisons for a range of fat-tailed solutions are shown in Figure~\ref{fig:fat}. The numerical computation of the exact solutions requires some care. We use the fact that they can be written as the density of \Levy-stable laws with a nonlinear rescaling (see~\cite{MP1}). A numerical method for computing these densities may be found in~\cite{Nolan}. For higher $\rho$, the error in the tails is negligible, showing that both the exact and computed density decay fast. However, the error near $s=0$ can be high (between $20\%$ and $30\%$ in the worst case observed), but the error decays rapidly with $s$ for  all $\rho\in(0, 1)$. This error is caused by the singularity near $s=0$ of the exact solutions $n_\rho$, $0 < \rho <1$. It is important to note however that the convergence of the scheme could be seen without a priori knowledge of the exact solutions. The initial  number of particles was $O(10^3)$, and the computation was terminated when the number of particles reached a maximal number $2 \times 10^7$ (fixed a priori). At each step of the dynamic renormalization, the number of particles must increase since the total number of particles is divergent for each of the exact solutions $n_\rho$. While our numerical scheme could be adapted to provide higher resolution (e.g. by incorporating special basis functions at $s=0$ and near $s=\infty$ to account for divergences), we have refrained from doing so, in order to demonstrate the robust convergence of the scheme used here.

\section{Acknowledgements}
One of the authors (I.G.K.) would like to remember here Stephen A. Orszag, who suggested this very problem as a challenge for equation free methods a decade ago. 

The authors also acknowledge support from the following funding agencies.
M.O.W. acknowledges support from NSF DMS 1204783.  I.G.K. acknowledges partial support from US-AFOSR through grant number FA9550-12-1-0332 and NSF grant CDSE 1310173. G.M. acknowledges partial support from NSF DMS 1411278.

\bibliographystyle{siam}      
\bibliography{lkmw}  
\end{document}